\documentclass[prb,aps,twocolumn,showpacs,home,floats]{revtex4}

\usepackage{graphics}
\begin{document}

\title{Hydrogen adsorption at RuO$_2$(110) }

\author{Qiang Sun, Karsten Reuter and Matthias Scheffler}

\address{Fritz-Haber-Institut der Max-Planck-Gesellschaft, 
Faradayweg 4-6,D-14195 Berlin-Dahlem, Germany}

\date{\today}

\begin{abstract}
The structural, vibrational, energetic and electronic properties of
hydrogen at the stoichiometric RuO${}_2$(110) termination are studied
using density functional theory. The oxide surface is found to 
stabilize both molecular and dissociated H$_2$. The most stable 
configuration in form of hydroxyl groups (monohydrides) at the 
undercoordinated O${}^{\rm br}$ surface anions is at low temperatures
accessed via a molecular state at the undercoordinated Ru${}^{\rm cus}$ 
atoms (dihydrogen) and a second precursor in form of a water-like 
species (dihydride) at the O${}^{\rm br}$ sites. This complex picture 
of the low-temperature dissociation kinetics of H${}_2$ at RuO${}_2$(110)
is in agreement with existing data from high-resolution energy-loss 
spectroscopy and temperature programmed desorption. Hydrogen adsorption at 
O${}^{\rm br}$ sites increases the reactivity of the neighboring 
Ru${}^{\rm cus}$ sites, which are believed to be the active sites in 
catalytic oxidation reactions.
\end{abstract}



\maketitle

\section{Introduction}
The widespread use of transition metal (TM) oxides, for example in applications 
for catalysis, electrochemistry, gas sensors, and corrosion/wear protection is 
an increasing source of motivation for fundamental research on this material 
class. An important goal in such studies, focusing on the surface functionality 
of oxides, is to establish atomic-scale insight into their surface structure and 
composition, as well as their interaction with gas phase species 
\cite{henrich94,noguera94}. Despite notable efforts secure knowledge is still 
rather scarce. This holds even for well-defined single-crystal surfaces under
the controlled conditions of ultra-high vacuum (UHV), mostly due to the 
structural complexity of oxides and to their often insulating nature which 
hampers the use of electron spectroscopy techniques \cite{woodruff94}.

With respect to these issues crystalline RuO${}_2$ represents a rather nice 
choice for a suitable benchmark model system. Not only is it one of the few 
metallic TM oxides and its rutile bulk structure of modest complexity 
\cite{sorantin92}, but with e.g. a reported high catalytic activity in oxidation 
reactions \cite{peden86,over00,over03} and being discussed as playing a 
sensitive role in Pt-Ru based direct methanol fuel cells \cite{rolison99}
it is also sufficiently interesting from an applied perspective. Especially the 
low energy RuO${}_2$(110) surface has recently received particular attention 
\cite{over00,over03,kim00,fan01,wang01,reuter02,reuter03a,reuter03b,reuter03c,
sun03a}. Focusing mostly on the CO oxidation reaction quite some detailed 
understanding (experimental as well as theoretical) on the fundamental 
interaction of O and CO with this model surface has emerged from these studies. 
From the potential interest for both catalytic and fuel cell applications it now 
appears natural to extend this knowledge also to the interaction of hydrogen
with RuO${}_2$(110). From a chemical point of view this ubiquitous gas phase 
species is expected to form strong bonds particularly with the oxygen anions at 
the oxide surface \cite{noguera94}. Intentionally or unnoticed hydrogen could 
therefore be present as surface species in a wide range of conditions, 
significantly influencing the functionality in the targeted application. In fact 
a noticeable effect of hydrogen contamination on the CO turnover numbers has 
been discussed in a recent experimental study \cite{zang00}.

On a microscopic level Wang {\em et al.} have provided detailed kinetic and 
vibrational data on the low-temperature hydrogen adsorption at RuO${}_2$(110) in 
UHV \cite{wang03}. Using high-resolution electron energy-loss spectroscopy 
(HREELS) and temperature programmed desorption (TPD) they identified both a 
molecular and a dissociated hydrogen state, the latter exhibiting vibrational 
properties of a water-like species with, however, a peculiarly blue-shifted 
scissor mode. Motivated by these specific results we set out to systematically 
investigate the structural, vibrational, energetic and electronic properties of 
hydrogen at the stoichiometric RuO${}_2$(110) surface using density-functional
theory (DFT) \cite{dreizler90}. The stability of molecular and dissociated 
hydrogen is first discussed at either of the two prominent adsorption sites 
exhibited by this oxide surface (Section IVA and IVB). Then higher coverages 
involving simultaneous occupation of both sites are addressed (Section IVC).
The detailed picture on the low-temperature dissociation kinetics of
H${}_2$ that emerges from the synergetic interplay between computations and 
experiments is intimately connected with the interesting property of 
RuO${}_2$(110) to simultaneously sustain both molecular and dissociated hydrogen 
states, as has already been described briefly in a preceding communication 
\cite{wang03}.

\section{Computational details}

The DFT calculations were performed within the Full-Potential Linear Augmented 
Plane Wave (FP-LAPW)  method \cite{blaha99,kohler96,petersen00} using the 
generalized gradient approximation (GGA) \cite{perdew96} for the 
exchange-correlation functional. The FP-LAPW basis-set parameters are as follows: 
R$_{MT}^{\rm Ru}$=1.8 bohr, R$_{MT}^{\rm O}$=1.0 bohr, R$_{MT}^{\rm H}$=0.6 
bohr, E$_{max}^{wf}$=20.25 Ry, E$_{max}^{pot}$=400 Ry, wave function expansion 
inside the muffin tins up to $l_{max}^{wf}$=12, and potential expansion up to 
$l_{max}^{pot}$=6. The Brillouin zone integration employed a $(4 \times 9 \times 
1)$ Monkhorst-Pack grid with 18 k-points in the irreducible Brillouin-zone (IBZ) 
for $(1 \times 1)$ surface unit-cells, and a $(4 \times 4 \times 1)$ grid with 8 
k-points in the IBZ for $(1 \times 2)$ cells. 

We stress that the very short O-H bonds represent a formidable challenge to 
electronic structure theory calculations. With respect to the FP-LAPW method 
employed this translates into the necessity to use rather small non-touching 
muffin-tin spheres. Then, the convergence behavior with respect to the 
interstitial plane wave cutoff is rather slow, requiring to use much higher 
E$_{max}^{wf}$ than the $\sim 17-19$\,Ry typical for late TM oxide studies 
involving muffin-tin sphere sizes only dictated by the O-metal bonds. 
Correspondingly, we tested this convergence by increasing the plane wave cutoff 
up to 36 Ry (!), as well as by employing denser k-meshes up to 36 k-points in 
the $(1 \times 1)$ IBZ. From these detailed tests (cf. the appendix), we  
conclude that the structural and vibrational properties of the systems 
addressed in the present study are well converged within 0.02\,{\AA} and 8\% 
respectively at the above stated E$_{max}^{wf}$=20.25 Ry, which correspondingly 
was chosen as our standard cutoff. Relative binding energy differences between 
configurations involving the same number of O and H atoms (e.g. when comparing 
tilted and untilted geometries) are similarly well converged to within 0.05\,eV 
at this cutoff. For absolute binding energies on the other hand, in particular 
when they involve strong O-H bonds like in hydroxyl groups, the chosen, already 
rather high cutoff is however still not enough. Where such numbers were required
within a 0.1\,eV/H atom numerical accuracy for our physical argument and are 
then listed in this manuscript, we correspondingly ran subsequent calculations
at a very (say for routine calculations still prohibitively) high cutoff 
of E$_{max}^{wf}$=30 Ry employing the structure relaxed before at 
E$_{max}^{wf}$=20.25 Ry. Further, we also checked on the uncertainty introduced 
by the use of the approximate exchange-correlation functional by performing a 
number of calculations employing also the local-density approximation (LDA) 
\cite{perdew92} and will comment on the differences between results obtained 
within LDA and GGA below.

The RuO$_2$(110) surface was modelled by a three trilayer O-(Ru${}_2$O${}_2$)-O
periodic slab as detailed before \cite{reuter02,reuter03a}, using a vacuum 
region of about 13\,{\AA} to decouple the interactions between neighboring slabs 
in the supercell geometry. All structures were fully relaxed by a damped Newton 
scheme until the residual forces on the atoms were less than 50\,meV/{\AA}, 
keeping only the atomic positions in the central trilayer at their fixed bulk 
positions. Test calculations employing 5 and 7 trilayer slabs and relaxing also 
deeper lying layers showed no significant structural changes beyond the topmost 
trilayer, neither was there an influence on the atomic surface geometries as 
obtained with the three trilayer calculations. At this point we would further 
like to emphasize that the structural relaxation allowed for any symmetry 
breaking at the surface. This was found to be crucial to obtain the correct 
energetics and structures, which often involve significant tilting of the 
surface groups.

For the calculations of the vibrational modes of the various surface species, 
the dynamical matrix was set up by displacing each of the involved surface atoms 
from their equilibrium positions in 0.04\,{\AA} steps. Anticipating a good 
decoupling of the vibrational modes due to the large mass difference between Ru 
and O/H, the positions of all atoms in the substrate below the adsorption site 
were kept fixed in these calculations. The normal modes were then obtained by 
subsequent diagonalization of the dynamic matrix.

\section{Bulk RuO$_2$, clean (110) surface, free H$_2$ and H$_2$O} 

Summarizing and extending the results of our earlier publications on RuO${}_2$ 
\cite{reuter02,reuter03a,reuter03b,reuter03c,sun03a} we first briefly describe 
the obtained properties of the bulk and the clean (110) surface, as well as free 
H${}_2$ and H${}_2$O molecules, as far as they are relevant for the 
understanding of the hydrogenated surface discussed below.

\begin{figure}
\scalebox{0.6}{\includegraphics{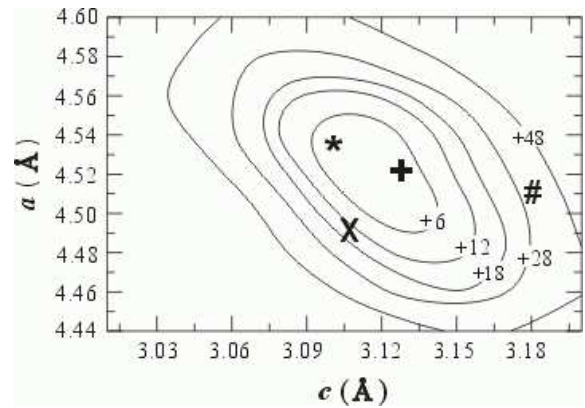}}
\caption{\label{fig1}
Computed GGA energy contours as a function of RuO${}_2$ bulk lattice parameters 
$a$ and $c$ (with optimized internal parameter $u$). The energy zero corresponds 
to the optimized values of $a = 4.52$\,{\AA} and $c = 3.13$\,{\AA} marked with 
`{\bf +}' and higher contour levels are labelled in the figure in meV per 
RuO${}_2$ formula unit. Additionally shown are the bulk lattice parameters as 
determined by experiments ( '${\bf *}$' $[$\onlinecite{kim00}$]$, '${\bf \sf 
x}$' $[$\onlinecite{huang82}$]$, '${\bf \#}$' $[$\onlinecite{atanasoska88}$]$ ).
The bulk lattice parameters obtained by an earlier DFT-GGA pseudopotential 
study are  $a = 4.65$\,{\AA} and $c = 3.23$\,{\AA} $[$\onlinecite{kim00}$]$, 
outside of the range shown in the figure.}
\end{figure}

\begin{table}
\caption{\label{tableI}
RuO${}_2$ lattice constants ($a$, $c$ and $u$) and bulk modulus ($B_0$) as 
determined within LDA and GGA, full-potential (FP) and pseudopotential (PP) 
calculations. Experimental values are from X-ray diffraction (XRD) and low-
energy electron diffraction (LEED) experiments.}
\begin{center}
\begin{tabular}{lllllr}
                  & $a$ (\AA) & $c$ (\AA)& $u$ & $B_0$ (GPa) \\ \hline \hline
This work (FP-LDA)          &  4.42  &  3.05  & 0.306  & 352 \\
This work (FP-GGA)          &  4.52  &  3.13  & 0.306  & 294 \\ \hline
$[$\onlinecite{glassford93}$]$ (PP-LDA) &  4.56  &  3.16  & 0.307  & 283 \\
$[$\onlinecite{kim00}$]$ (PP-GGA)       &  4.65  &  3.23  & 0.305  & --  \\ 
\hline
$[$\onlinecite{boman70}$]$ (XRD)        &  4.492 &  3.106 & 0.306  & 270 \\
$[$\onlinecite{huang82}$]$ (XRD)        &  4.491 &  3.106 &  --    & --  \\ 
\hline
$[$\onlinecite{atanasoska88}$]$ (LEED)  &  4.51  &  3.18  &  --    & --  \\
$[$\onlinecite{kim00}$]$ (LEED)         &  4.51  &  3.23  &  --    & --  \\
\end{tabular}
\end{center}
\end{table}

RuO${}_2$ crystallizes in the rutile structure, described by lattice parameters 
$a$ and $c$, as well as one internal degree of freedom $u$ specifying the 
positions of the O anions within the bulk unit-cell \cite{sorantin92}. In order 
to determine the bulk equilibrium lattice constants we scanned a grid in $a$ and 
$c$ near the experimental values with 0.02\,{\AA} and 0.015\,{\AA} steps 
respectively at a high interstitial plane wave cutoff of 30 Ry. At each grid 
point the internal parameter $u$ was further optimized minimizing the
computed forces on the O atoms in the unit-cell. The resulting energy landscape 
is shown in Fig. \ref{fig1}, while Table \ref{tableI} lists the determined bulk 
lattice parameters within the LDA and GGA. Comparing with available bulk 
diffraction studies \cite{boman70,huang82} we obtain very good agreement within
the GGA. Interestingly, two earlier DFT pesudopotential calculations (LDA 
\cite{glassford93} and GGA \cite{kim00}) yield larger lattice parameters that 
deviate by more than 0.1{\AA} from our LDA and GGA values, respectively.
We particularly checked on the GGA value by Kim {\em et al.}, but find their 
geometry to be about 0.4\,eV per RuO${}_2$ formula unit higher in energy than 
our ground state geometry. Tentatively, we take this 0.1\,{\AA} and 0.4\,eV as a 
rough estimate of the structural and energetic error introduced by the 
pseudopotential approximation when describing late transition metal (TM) oxide 
compounds. Finally, we notice that the $c$ lattice parameter measured in two 
independent low-energy electron diffraction (LEED) experiments 
\cite{kim00,atanasoska88} deviates from the bulk diffraction results
by about 0.1{\AA}. In both LEED experiments thin RuO${}_2$(110) films grown on 
Ru(0001) were investigated, where the $c$ lattice parameter corresponds to the
size of the surface-unit cell in the [001] direction.  Apparently, a thin film 
(with an incommensurable structure to that of the metal substrate) prefers a 
slightly different geometry than a bulk crystal.

\begin{figure}
\scalebox{0.6}{\includegraphics{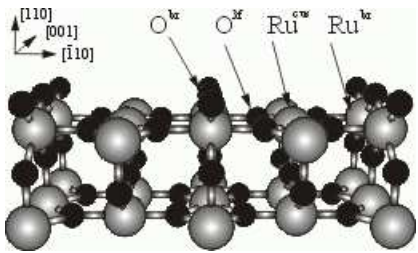}}
\caption{Side view of the stoichiometric RuO${}_2$(110) surface termination 
explaining the location of the two prominent adsorption sites corresponding to 
under-coordinated surface atoms: bridging oxygen O${}^{\rm br}$ and 
coordinatively unsaturated (cus) ruthenium Ru${}^{\rm cus}$ (Ru = light, large 
spheres; O = dark, small spheres).}
\label{fig2}
\end{figure}

Depending on the oxygen-content in the surrounding gas phase, either a 
stoichiometric or an oxygen-rich termination is stabilized at the RuO${}_2$(110) 
surface \cite{reuter02,sun03a}. Figure \ref{fig2} shows the surface geometry of 
the stoichiometric termination exhibiting two kinds of surface species, the 
nearest neighbor shell of which has been reduced by the creation of the surface: 
A twofold coordinated bridging oxygen O${}^{\rm br}$ (threefold coordinated in 
the bulk) and a fivefold coordinated ruthenium atom Ru${}^{\rm cus}$ (sixfold 
coordinated in the bulk). The only other surface species present, O${}^{\rm 
3f}$, still maintains its bulk-like threefold coordination to in-plane Ru atoms. 
The oxygen-rich termination differs from this geometry only by extra oxygen 
atoms adsorbed on-top of the Ru${}^{\rm cus}$ atoms. In the following we will 
focus exclusively on the interaction of hydrogen with the stoichiometric 
termination, attempting to make contact with the existing data from ultra-high 
vacuum (UHV) experiments \cite{wang03}, where the stoichiometric termination 
is the standard surface produced after a high-temperature anneal to 600\,K
\cite{over00,kim00,fan01,wang01}. The discussion of hydrogen interaction with 
the oxygen-rich termination is deferred to a consecutive publication.

Due to the cutting of bonds the surface atoms relax. Most prominently we find 
that the O${}^{\rm br}$ atoms move inwards, reducing their bond length to the 
underlying Ru atoms (henceforth denoted Ru${}^{\rm br}$) to 1.91\,{\AA} (bulk: 
1.99\,{\AA}), and, correspondingly, the inter-layer distance is reduced to 
1.09\,{\AA} (12\% smaller than bulk). These findings are in very good agreement 
with a recent LEED study which determined a bond length of 1.94\,{\AA} and a 
first layer contraction of -13\% at RuO${}_2$(110) \cite{kim00}, and they are  
similar to the geometric changes observed at the isostructural TiO${}_2$(110) 
surface \cite{henrich81}. Although not as pronounced also the under-coordinated 
Ru${}^{\rm cus}$ atoms relax a bit inwards, thereby inducing a buckling 
within the first trilayer plane of 0.16\,{\AA} (LEED: 0.18\,{\AA}). Contrary to 
other transition metal oxide surfaces like e.g. Al${}_2$O${}_3$(0001) 
\cite{wang00} the structural relaxations are therefore rather small, and in 
particular damp away rapidly: No significant deviations from the bulk-like 
positions are found for atoms below the first trilayer, neither in LEED, nor in 
our thicker slab test calculations.

\begin{table}
\caption{\label{tableII}
Structure parameters (bond length $d$ and angle), binding energy $E_b$, and 
vibrational frequencies $\nu$ for gas phase H${}_2$ and H${}_2$O. Compared are 
the computed values for LDA and GGA (at two different LAPW cutoffs) 
with the corresponding experimental data (with zero-point energy removed).}
\begin{center}
\begin{tabular}{l | rr | rr | r}
                      &\multicolumn{2}{c|}{LDA} & \multicolumn{2}{c|}{GGA} & 
Exp. \\ 
                      & 20.25\,Ry & 30\,Ry & 20.25\,Ry &\,30 Ry & \\ \hline
H$_2$  & & & &\\
$d_{\rm H-H}$ (\AA)   & 0.77 & 0.77 & 0.75 & 0.75 & 0.74 
$[$\onlinecite{crc00}$]$ \\
$E_b$ (eV)            & 4.79 & 4.87 & 4.45 & 4.56 & 4.73 
$[$\onlinecite{crc00}$]$ \\
$\nu_{\rm stretch}$ (meV)    &  --  &  --  & 538  &  --  & 546  
$[$\onlinecite{thiel78}$]$ \\
&&&&&\\
H$_2$O & & & &\\
$d_{\rm O-H}$ (\AA)   & 0.98  & 0.98  & 0.97  & 0.97   & 0.96 
$[$\onlinecite{thiel78}$]$ \\
$\angle_{\rm HOH}$ (deg) & 102   & 102   & 103   & 103    & 104.5 
$[$\onlinecite{thiel78}$]$ \\
$E_b$ (eV)            & 11.08 & 11.49  & 10.04  & 10.40  & 10.06 
$[$\onlinecite{nbs65}$]$ \\
$\nu_{\rm sym}$ (meV) &  --   & --    & 424   & --     & 454  
$[$\onlinecite{thiel78}$]$ \\
$\nu_{\rm asym}$ (meV)&  --   & --    & 435   & --     & 466  
$[$\onlinecite{thiel78}$]$ \\
$\nu_{\rm scissor}$ (meV)& -- & --    & 189   & --     & 198  
$[$\onlinecite{thiel78}$]$ \\ 
\end{tabular}
\end{center}
\end{table}

Finally, we summarize in Table \ref{tableII} the computed binding energies,
bond lengths and frequencies for gas phase H${}_2$ and H${}_2$O. For both 
molecular and atomic calculations we employed the same muffin-tin spheres as 
detailed before for the slab calculations, and allowed for non-spherical 
densities by reducing the symmetry. The total energies for the isolated, 
spin-polarized atoms are obtained by adding to the total energy value from a 
non-spin-polarized FP-LAPW calculation a constant spin-polarization energy of 
1.52\,eV (O) and 1.10\,eV (H) taken from a relativistic atomic DFT calculation 
\cite{blaha99}. With respect to structural and vibrational properties we obtain 
very good agreement with the experimental data, as well as with previous DFT 
studies \cite{clementi90,finocchi01,hammer95}. The slow convergence of the
O-H bond energetics already described in the preceding section exists
similarly for the free molecules, which is why we list the computed values both 
at the routine 20.25\,Ry and at the very high LAPW cutoff of 30\,Ry (used 
lateron to obtain quantitative binding energies as discussed above). With 
respect to the latter cutoff we assess the numerical convergence of the 
gas phase binding energies to be within 0.2\,eV. This points then at a sizable
error (i.e. overbinding) compared to the experimental values when oxygen
is involved even within the GGA, which is a well known result \cite{perdew96}. 
Although some error cancellation occurs in the computation of binding energies
at surfaces (as illustrated in the appendix), we conclude that a cautious 
reasoning is necessary when judging on the endo- or exothermicity of adsorption 
with respect to the gas phase molecules. To this extent most of our physical 
arguments will instead rather be based on binding energy {\em differences}, 
which are more accurately described ($\pm 0.05$\,eV, cf. the appendix).

\section{The hydrogenated RuO${}_2$(110) surface}

In the following sections we systematically discuss the hydrogen interaction 
with the various surface species present at the stoichiometric RuO${}_2$(110) 
termination. The energetics will be described with respect to the aforedescribed 
free H${}_2$ molecule, where a negative binding energy denotes endothermicity 
with respect to the gas phase species and the bare surface, i.e. a metastable 
situation. As we find the interaction with the threefold coordinated in-plane 
O${}^{\rm 3f}$ to be energetically very unfavorable (even the formation of a 
monohydride is endothermal by $\approx-0.3$\,eV/H atom), the discussion will 
concentrate on the two under-coordinated surface species, i.e., Ru${}^{\rm cus}$ 
and O${}^{\rm br}$. At first we discuss hydrogen first to be present at one of 
these sites only (Section IVA and IVB). And then we consider simultaneous 
adsorption at both sites (Section IVC). Consistent with our previous 
publications we will employ a short hand notation to characterize the 
manyfold of studied geometries, indicating first the occupancy of the bridge 
site and then of the cus site, e.g. (OH)${}^{\rm br}$/H${}_2^{\rm cus}$ 
for a configuration with an OH-group at the bridge site and a H${}_2$ 
molecule at the cus site. 

Lateral interactions between functional groups at directly neighboring cus and 
bridge sites are implicitly contained within our calculations employing 
$(1 \times 1)$ surface unit-cells. With these cells only integer multiples of 
one monolayer (ML) hydrogen coverage can be studied (1 ML defined as monoatomic 
occupation of all sites of one type (br or cus)). Further reaching lateral 
interactions, e.g. towards a moiety at the same site type either along 
$[\bar{1}10]$ (at a distance of 6.4\,{\AA}) or along $[001]$ (at a distance of 
3.1\,{\AA}, see Fig. \ref{fig3}) could lead to the formation of more dilute 
superstructures with fractional MLs hydrogen coverage. From a systematic study 
of oxygen adsorption at RuO${}_2$(110) we find such lateral interactions to be 
rather small at this rather open oxide surface \cite{reuter03d} and don't expect 
this to be significantly different for hydrogen, in particular along the longer 
$[\bar{1}10]$ direction of the surface unit-cell, where the next-nearest
site would be 6.4\,{\AA} away. Correspondingly, we only test for lateral
interactions in selected configurations employing larger $(1 \times 2)$ 
cells, allowing us to model structures where then only every second site 
is occupied along the $[001]$ direction (see e.g. Fig. \ref{fig7} below).

\subsection{Hydrogen at Ru${}^{\rm cus}$}

\begin{figure}
\scalebox{0.6}{\includegraphics{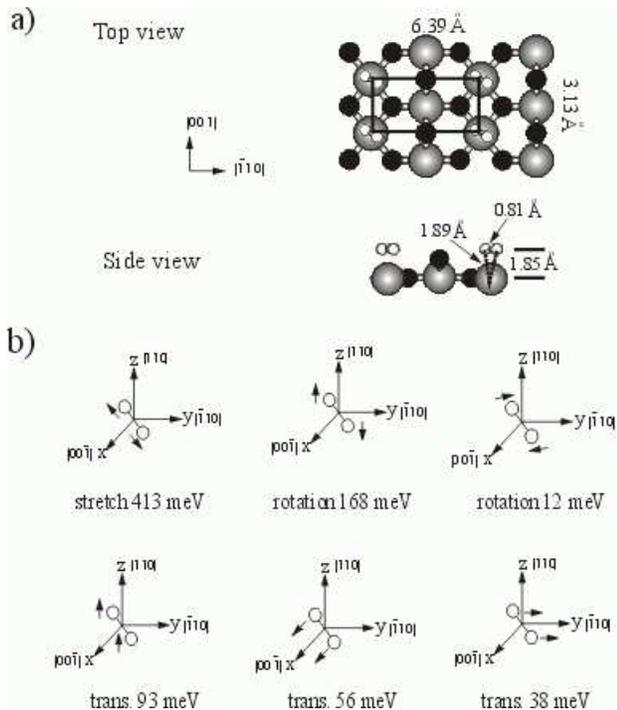}}
\caption{a) Top and side view of the adsorption geometry of 1\,ML
H${}_2$ at Ru$^{\rm cus}$ (Ru = light, large spheres; O = dark, small
spheres; H = white, small spheres). Additionally shown is the size of the $(1 
\times 1)$ surface unit-cell. Note that the azimuthal direction of the plotted 
H$_2$ is arbitrary, as it is freely rotating. b) Six vibrational modes of the 
adsorbed H${}_2$ molecule. The directions and lengths of the arrows indicate 
approximately the directions and relative magnitudes of the displacements of the
atoms.}
\label{fig3}
\end{figure}

We start with the molecular adsorption of H${}_2$ at the under-coordinated 
Ru${}^{\rm cus}$ site (\,O${}^{\rm br}$/H${}_2^{\rm cus}$\,). This analysis was 
motivated by the aforementioned recent UHV HREELS experiments that attributed a 
weak peak at 367\,meV to the stretch mode of a molecular hydrogen species at the 
surface with a TPD activation energy of about 0.3\,eV \cite{wang03}. Relaxing a 
H${}_2$ molecule in $(1 \times 1)$ cells from a height at about 2\,{\AA} atop 
the cus sites our calculations indeed find such a (meta)stable species with a 
computed binding energy of $+0.32$\,eV/H${}_2$ with respect to molecular 
H${}_2$. The resulting adsorption geometry is shown in Fig. \ref{fig3} together 
with the calculated vibrational modes. At a height of 1.85\,{\AA} the H${}_2$ 
molecule lies parallel to the surface above the cus sites (side-on 
configuration). Interestingly, we find almost no corrugation of the potential 
energy surface with respect to an azimuthal rotation of the flat-lying H${}_2$ 
molecule: The optimal bond orientation about 30$^{\circ}$ degree from the 
[$\bar{1}$10] direction is only by insignificant 3\,meV more stable than any 
other orientation, i.e. the H${}_2$ behaves essentially like a free-rotator 
(helicopter mode), as also reflected by the very low rotational vibration 
frequency of 12\,meV, cf. Fig. \ref{fig3}. The similarly low in-plane 
translational modes further point at a rather expressed delocalization of the 
H${}_2$ molecule parallel to the surface, particularly in the [$\bar{1}$10] 
direction, i.e. approaching the neighboring bridging oxygens. Despite this the 
H${}_2$ bond length is with 0.81\,{\AA} noticeably stretched compared to the 
free gas phase molecule (0.75\,{\AA}). This is consistent with the rather high 
binding energy and the significantly decreased stretch vibration (413\,meV 
compared to 538\,meV in the gas phase). 

These properties reflecting a moderately strong interaction are very similar to 
molecular H${}_2$ at late transition metal surfaces (like e.g. Pd(100) 
\cite{wilke96}) if adsorption is restricted to take place at the on-top site. 
Of course, at the latter surfaces the on-top site is only a local ``constrained'' 
minimum and H${}_2$ would dissociate towards higher-coordinated hollow sites. 
The different geometry of RuO${}_2$(110) doesn't offer such sites, thus 
stabilizing the molecular adsorption at Ru${}^{\rm cus}$. In this respect we 
attribute the non-dissociative interaction of H${}_2$ with these sites more to a 
geometry effect, rather than to an electronic effect, i.e. compared with the 
weak physisorption of H${}_2$ at noble metal surfaces like Ag(111) 
\cite{gruyters94}.

\begin{figure}
\scalebox{0.6}{\includegraphics{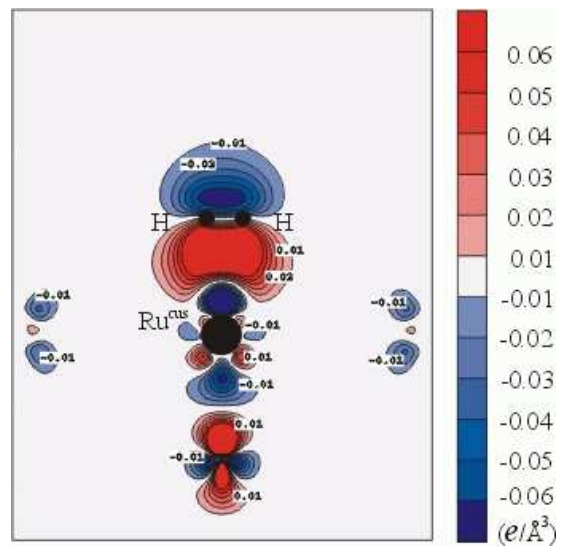}}
\caption{Difference density plot for H${}_2$ adsorbed at Ru$^{\rm cus}$. The 
contour plot depicts the plane parallel to the H${}_2$ molecular axis and normal 
to the surface. Areas of electron accumulation and depletion have positive and 
negative signs respectively, contour lines are drawn at 0.01\,$e$/{\AA}$^{3}$ 
intervals.}
\label{fig4}
\end{figure}

The relatively strong interaction (electron polarization and bond formation) is 
also obvious  in the ``difference density'' plot shown in Fig. \ref{fig4}. The 
``difference density" is obtained by subtracting from the electron density of the 
adsorbate system that of the clean surface and that of the free molecule (the 
latter two with the same interatomic distances as found for the adsorpbate 
system) \cite{scheffler00}. A rather strong polarization of the flat-lying
H${}_2$ is apparent, with electron density accumulation on the substrate side of 
the molecule and depletion on the vacuum side. Due to the interaction with the 
H${}_2$ the back-bond of the Ru$^{\rm cus}$ atom to the underlying O substrate 
atom is slightly weakened, the Ru atom moves 0.03\,{\AA} upwards and thereby
reduces the clean-surface-buckling in the topmost trilayer.

The structure of the difference density plot suggests the major interaction in 
the occupied states to be due to a hybridization of Ru-$d_{z^2}$ and 
H${}_2$-$\sigma$ orbitals. Analyzing the computed local density of states
we indeed find the bonding governed by this and a smaller hybridization
of the Ru-$d_{xz}$/$d_{yz}$ with the H${}_2$-$\sigma^*$ orbitals. The bonding
of the H${}_2$ molecule to the cus site can therefore be understood
within the familiar donation/back-donation picture \cite{scheffler00},
where the hybridization with the H${}_2$-$\sigma$ orbital causes a small
H${}_2$$\rightarrow$Ru charge transfer that is counteracted by some 
back-donation of electronic charge from the metal to the H${}_2$-$\sigma^*$ 
level, strengthening the coupling to the substrate while weakening and 
elongating the H-H bond.

Contrary to the situation at most more reactive TM surfaces this back-donation 
is however not strong enough to completely dissociate the H${}_2$ molecule. This 
is supplemented by the surprising result that atomic hydrogen is not stable at 
the Ru${}^{\rm cus}$ site: We compute only an endothermal binding energy of $-
0.33$\,eV/H atom with respect to 1/2 H${}_2$ at 1\,ML H-coverage, i.e. in $(1 
\times 1)$ unit-cells. Checking whether H bonding might become more favorable in 
more dilute superstructures we also employed $(1 \times 2)$ cells to model a 
0.5\,ML H-coverage with H only at every second site along the $[001]$ direction. 
With -0.20\,eV/H atom with respect to 1/2 H${}_2$ the binding energy is still 
endothermal and only slightly changed, reflecting the small lateral interactions 
at the RuO${}_2$(110) surface. Expecting no further changes in binding energy
for even more dilute H-phases we therefore conclude that only molecular
hydrogen may be stabilized at the cus sites. 

Forming the basis for e.g. relations between heterogeneous and homogeneous 
catalysis it is interesting to compare these findings for the Ru atom at the 
surface of an oxide with the hydrogen bonding to TM atoms in other frameworks 
like e.g. at the surface of metals or in a TM complex. As already mentioned, 
coupling of hydrogen to TM surfaces is generally associated with the 
dissociation of the ligand \cite{christmann88}. The observation of 
non-dissociative chemisorption of H${}_2$ has so far been restricted to a few 
exceptional cases \cite{martensson86,schmidt01}, mostly connected with a prior 
saturation of the most reactive sites at the surface with atomic hydrogen. 
Concomitantly, we compute e.g. the bonding of atomic H at Ru(0001) to be 
exothermic at least up to 1\,ML coverage. Molecular precursors have more been 
identified at noble surfaces like Ag(111) \cite{gruyters94}, yet then 
physisorbed and certainly not exhibiting such a strong activation of the H${}_2$ 
bond as expressed at RuO${}_2$(110) with the significant bond elongation and 
downshift of the stretch frequency. These findings resemble much more the data 
from organometallic complexes: For so-called $\eta^2$-H${}_2$ (dihydrogen) 
single metal atom complexes, in which the H-H bond remains intact 
\cite{kubas84}, TM-H${}_2$ bond energies in the range of 0.1-0.3\,eV/H${}_2$ are 
estimated \cite{crabtree86}, significant red-shifts of the stretch frequency 
\cite{crabtree86,kubas91} and bond elongation up to 0.9\,{\AA} are reported 
\cite{zilm86,sluys90}. In fact, neutron scattering experiments furthermore 
indicate rapid rotation of $\eta^2$-H${}_2$ ligands with an activation energy of 
less than about 10\,meV \cite{zilm86}, just as we find for the molecular 
hydrogen at Ru${}^{\rm cus}$. Even the donation/back-donation bonding model is 
analogously discussed for the TM complexes \cite{brunner96}. Yet, there (just 
like at TM surfaces) the back-donation may also break the H-H bond, and often 
H${}_2$- as well as H-ligands are attached to the same metal center and can even 
exhibit continuous changes between both configurations 
\cite{crabtree86,kubas91}. At RuO${}_2$(110), in contrast, the bonding to the 
surrounding oxide apparently depletes the electron density at the 
under-coordinated Ru${}^{\rm cus}$ atom already in such a way, that 
the back-bonding only activates the H${}_2$ bond, but may no longer break it. 

\subsection{Hydrogen at O${}^{\rm br}$}

\begin{figure}
\scalebox{0.6}{\includegraphics{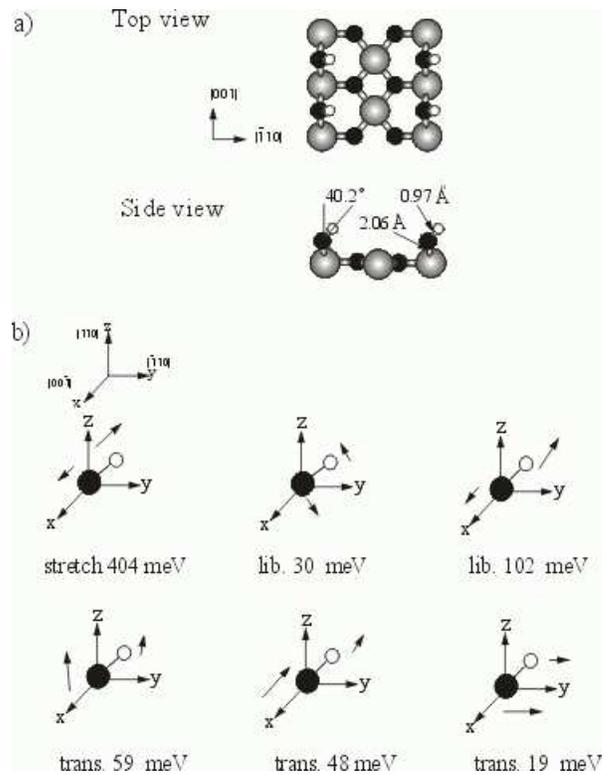}}
\caption{a) Top and side view of the adsorption geometry of 1\,ML H at O$^{\rm
br}$, computed in $(1 \times 1)$ unit-cells. b) Six vibrational modes of the
formed hydroxyl group.}
\label{fig5}
\end{figure}

\begin{figure}
\scalebox{0.6}{\includegraphics{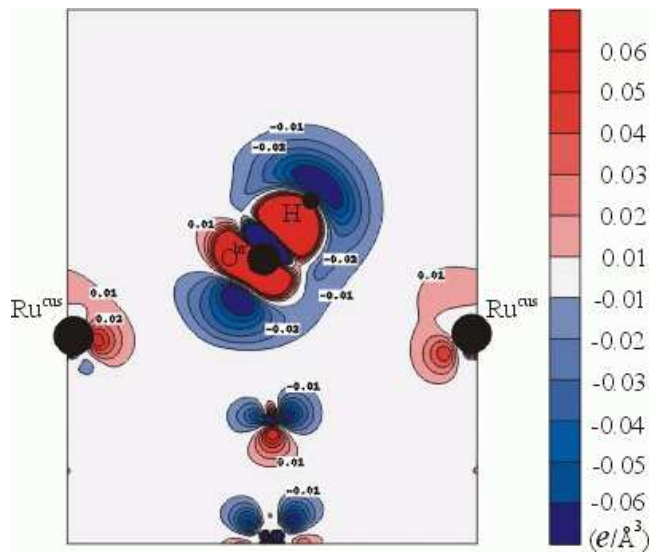}}
\caption{Difference density plot of a hydroxyl group at the bridge sites
(OH)${}^{\rm br}$/$-$. The contour plot depicts the plane parallel to
$[\bar{1}10]$ and normal to the surface. Areas of electron accumulation
and depletion have positive and negative signs respectively, contour lines are
drawn at 0.01\,$e$/{\AA}$^{3}$ intervals.}
\label{fig6}
\end{figure}

For H$_2$ molecules approaching the RuO$_2$ surface at the under-coordinated 
O${}^{\rm br}$ atoms we find that they will either slide towards the Ru${}^{\rm 
cus}$ sites or that they are repelled into the gas phase. In no case (i.e., 
testing many H$_2$ orientations, as well as lower coverages in a $(1 \times 2)$ 
cell) did we observe spontaneous dissociation above the O${}^{\rm br}$ site.
This leads us to conclude that molecular adsorption primarily takes place over 
the aforediscussed cus sites. Atomic hydrogen, on the other hand, binds
at the O${}^{\rm br}$ sites, i.e., forming a surface hydroxyl group. The 
optimized geometry of the 1\,ML (OH)${}^{\rm br}$/$-$ phase computed in $(1 
\times 1)$ cells is shown in Fig. \ref{fig5}. The strong binding within the 
OH-group weakens the bond to the underlying Ru${}^{\rm br}$ substrate atoms and 
elongates it significantly from 1.91\,{\AA} to 2.06\,{\AA}. The surface OH-group 
itself has a bond length of 0.97\,{\AA}, nearly identical to the O-H distance in 
H$_2$O, and is inclined towards the [$\bar{1}$10] direction with a 40$^{\circ}$ 
angle with respect to the surface normal. The computed binding energy is 
+0.89\,eV/H atom stronger than in molecular H$_2$, and the energy gain by the 
tilting is 0.1\,eV/H atom compared to the higher-symmetry, up-right 
configuration. Again, both the computed stretch frequency, as well as the 
binding energy are found to be in good agreement with the recent HREELS and TPD 
data from the hydroxylated surface \cite{wang03}, and the strong binding is also 
nicely visible in the difference density plot shown in Fig. \ref{fig6}.

\begin{figure}
\scalebox{0.6}{\includegraphics{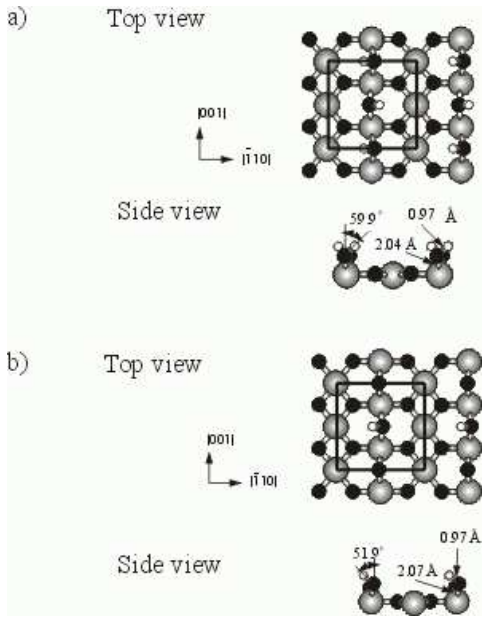}}
\caption{a) Top and side view of the adsorption geometry of 1\,ML
H at O$^{\rm br}$, computed in $(1 \times 2)$ unit-cells with alternating
tilting. b) Top and side view of the adsorption geometry of 0.5\,ML
H at O${}^{\rm br}$.}
\label{fig7}
\end{figure}

With only a 0.1\,eV energy difference between tilted and upright position, 
the hydroxyl groups will at finite temperatures frequently swing from one 
orientation to the other. Checking whether this may occur in concerted wave-like 
motions along a chain of bridge sites, we repeated the calculation with 
1\,ML coverage, but now in a $(1 \times 2)$ cell with each OH-group 
alternatingly tilted in one or the other direction as shown in Fig. \ref{fig7}a. 
The only structural difference obtained is a somewhat larger tilt angle of about 
60${}^{\circ}$ degree, while bond length and binding energy remain to within 
0.01\,eV/H atom virtually unchanged. Testing for further reaching lateral 
interactions by decreasing the total coverage to 0.5\,ML as shown in Fig. 
\ref{fig7}b, we again find the binding energy within 0.02\,eV/H atom degenerate 
to the previous two cases, leading us to conclude that each OH-group tilts and 
swings essentially independent of the others, and of the total coverage which 
may thus easily reach the full 1\,ML with each O${}^{\rm br}$ atom hydroxylated.

\begin{figure}
\scalebox{0.6}{\includegraphics{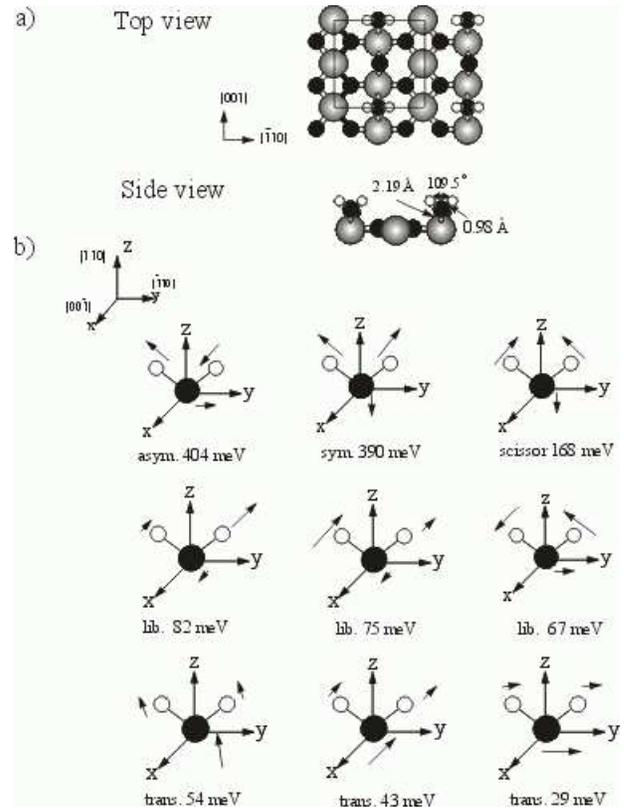}}
\caption{a) Top and side view of the adsorption geometry of 0.5\,ML water-like 
species at the bridge sites, oriented along the $[\bar{1}10]$ direction. b) 
Corresponding nine vibrational modes.}
\label{fig8}
\end{figure}

\begin{figure}
\scalebox{0.6}{\includegraphics{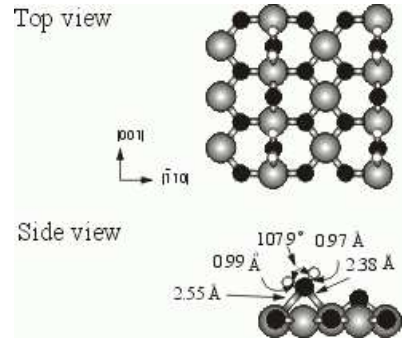}}
\caption{Top and side view of the adsorption geometry of 0.5\,ML
water-like species at the bridge sites, oriented along the $[001]$ direction.}
\label{fig9}
\end{figure}

The low temperature HREELS and TPD experiments by Wang {\em et al.} 
\cite{wang03} show that the coverage of dissociated hydrogen can be increased 
above 1ML. There is evidence for a hydrogen state at O${}^{\rm br}$, that 
exhibits frequencies resembling a water-like H-O-H configuration, i.e. with two 
H atoms per bridge site (dihydride). Surprisingly, the measured scissor mode for
this functional group is with 231\,meV higher than the one of gas phase water 
(exp: 198\,meV, calc: 189\,meV, see Table \ref{tableII}), whereas intuitively a 
red-shift would be expected: Considering that the molecule-surface interaction
weakens the intra-molecular bonds and normally widens the H-O-H angle, both 
these effects would rather tend to make the bending mode softer 
\cite{christmann88}. Trying to address this puzzling finding we computed such 
water-like configurations at the bridge sites first at 1\,ML H${}_2$O-coverage 
in $(1 \times 1)$ cells and with the water-axis oriented once along $[001]$ and 
once along $[\bar{1}10]$. Both configurations exhibit only a very low stability
and can thus not account for the experimental water-like species with a 
desorption temperature around 300\,K \cite{wang03}. We therefore proceeded to 
relax the same two configurations at lower coverages in $(1 \times 2)$ cells 
with water-like species now only at every second bridge site as shown in Figs. 
\ref{fig8} and \ref{fig9}. While the model with orientation along the [001] axis 
is still slightly endothermal, the binding energy of the other orientation shown 
in Fig. \ref{fig8} finally turns out already at least exothermal by 
+0.48\,eV/H${}_2$. Still, as almost expected the calculated scissor mode for 
this configuration (just as much as the one of the other three models) is with 
168\,meV significantly lower than the one of free water, and thus in strong 
disagreement with the experimental data.

\begin{figure}
\scalebox{0.6}{\includegraphics{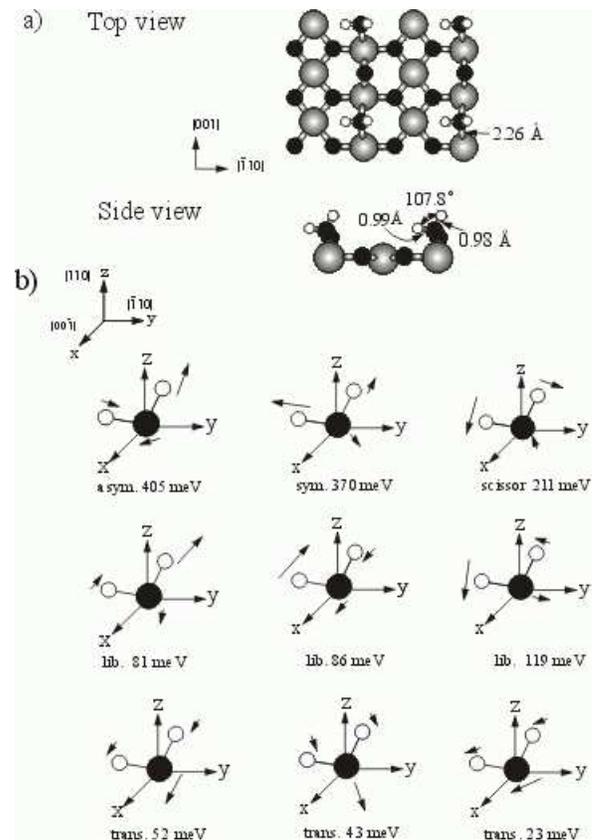}}
\caption{a) Top and side view of the adsorption geometry of 0.5\,ML
water-like species at the bridge sites, oriented asymmetrically
along the $[\bar{1}10]$ direction. b) Corresponding nine vibrational 
modes.}
\label{fig10}
\end{figure}

Recalling the strong tilting of the mono-hydride group at O${}^{\rm br}$
shown in Fig. \ref{fig5} we then tested to similarly tilt the whole water-like 
species of the last most favorable model and ended up with the geometry shown in 
Fig. \ref{fig10}, that is apparently separated from the up-right configuration 
by a sizable energy barrier (i.e. neither configuration relaxes automatically 
into the other one). The tilt not only significantly increases the binding 
energy of this configuration by 0.08\,eV to +0.56\,eV/H${}_2$, but also the 
computed scissor mode is with 211\,meV 12\% higher than the one of a free water 
molecule (189\,meV, see Table \ref{tableII}). Also the other computed 
vibrational modes listed in Fig. \ref{fig10}b are now in reassuring agreement 
with the experimental HREELS data (exp: stretch 436\,meV, scissor 231\,meV, 
libration 110\,meV 76\,meV, translation 59\,meV 28\,meV) \cite{wang03}. 

\begin{figure}
\scalebox{0.6}{\includegraphics{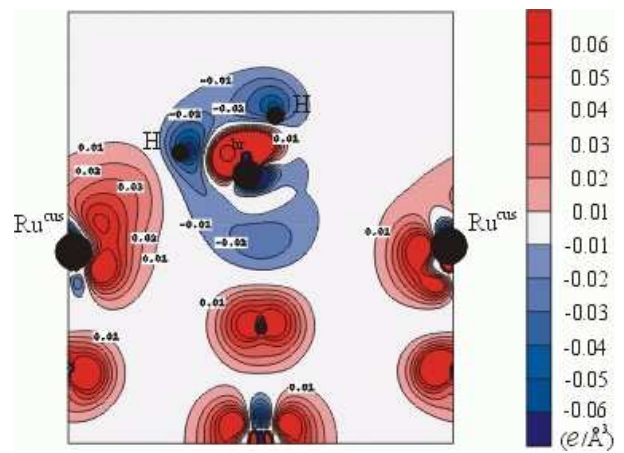}}
\caption{Difference density plot of the asymmetric water-like species of Fig. 
\ref{fig10} at the bridge sites. The contour plot depicts the plane parallel to 
$[\bar{1}10]$ and normal to the surface. Areas of electron accumulation and 
depletion have positive and negative signs respectively, contour lines are drawn 
at 0.01\,$e$/{\AA}$^{3}$ intervals.}
\label{fig11}
\end{figure}

The question remains why this tilted geometry gives rise to the 
counterintuitively blue-shifted scissor mode. Inspecting the relaxed geometry 
displayed in Fig. \ref{fig10} in more detail we find the water-like H-O-H bond 
angle with 108${}^{\circ}$ slightly increased compared to free water 
(calculation: 103${}^{\circ}$, see Table \ref{tableII}), lending more towards 
the argument favoring a weakening of the bending mode. However, this angle 
together with the overall tilting of the whole functional group causes the lower 
OH-bond of the water-like species to end up almost parallel to the surface (only 
8${}^{\circ}$ to the surface plane), bringing the terminal H atom closer to the 
neighboring Ru${}^{\rm cus}$ atom (2.72\,{\AA} compared to 3.19\,{\AA} in the 
upright model shown in Fig. \ref{fig8}). This suggests a possible interaction 
between these two species that is indeed verified by inspecting the computed
difference density plot shown in Fig. \ref{fig11}. Although small, the apparent 
hydrogen bridge bonding possible in this tilted configuration hinders any 
movement away from Ru${}^{\rm cus}$ parallel to the $[\bar{1}10]$ direction and 
thus naturally stiffens not only the scissor mode, but also the corresponding 
librational mode (calc: 119\,meV, exp: 110\,meV, compared to ``typical'' values 
around 80\,meV, cf. Figs. \ref{fig10} and \ref{fig8}).

This intricate coupling between neighboring bridge and cus sites fits nicely 
into the picture of the low temperature dissociation kinetics of H${}_2$ at 
RuO${}_2$(110) that emerges from the data presented in Sections IVA and IVB, as 
well as from the detailed UHV-HREELS experiments by Wang {\em et al.} \cite{wang03}. 
Although the thermodynamic ground state for hydrogen at this stoichiometric 
surface is given by the strongly bound hydroxyl-groups at the bridge sites 
(binding energy +0.89\,eV/H\,atom) \cite{sun03a}, direct dissociation over
these sites is apparently inhibited at least at low temperatures by an energy 
barrier. Thus, we tend to conclude on the following scenario: When H$_2$ 
interacts with RuO$_2$(110), it is at first bound in molecular form at the cus 
sites (binding energy +0.16\,eV/H\,atom). Although the H${}_2$ bond is weakened 
in this process, the cus sites can not induce its complete cleavage. This is 
instead achieved via the water-like metastable configuration at the bridge sites 
(binding energy +0.28\,eV/H\,atom) that may be accessed from the molecular 
H${}_2$ state at the cus sites (as indicated by the low translational mode in 
$[\bar{1}10]$ direction, cf. Fig. \ref{fig3}). From there the hydroxyl groups 
are finally formed, presumably by activated hydrogen diffusion. This 
interpretation of H${}_2^{\rm cus}$ as a necessary precursor state to 
dissociation is also supported by new low temperature HREELS and TDS experiments 
which report a suppressed population of the water-like species at bridge if the 
cus sites are first blocked by CO molecules \cite{wang03b}. Only if the water-
like species are allowed to form, the hydroxyl groups result from moderate 
heating to 350\,K \cite{wang03}.

\subsection{Hydrogen at both sites: Higher coverages}

Having discussed the lower coverage adsorption up to 1\,H${}_2$-ML for both 
sites separately, we now proceed to higher total coverages involving hydrogen at 
both sites. Also with various hydrogen functional groups present at the bridge 
sites we still find atomic hydrogen at cus to be always unstable. 
Correspondingly we restrict our detailed discussion to molecular H${}_2$ at cus 
and different hydrogen populations at the bridge sites. Starting with
the water-like species at bridge, the $(1 \times 1)$ configuration (H${}_2$O)${}^{\rm 
br}$/H${}_2^{\rm cus}$ corresponding to a total coverage of 2\,H${}_2$-ML is 
computed to be endothermal, reflecting presumably already an oversaturation of 
the surface with hydrogen. This improves in more dilute $(1 \times 2)$ 
configurations with (H${}_2$O)${}^{\rm br}$ and H${}_2^{\rm cus}$ occupying only 
every other site. The average binding energy for both a checkerboard arrangement 
and with both species occupying directly neighboring sites turns then out 
exothermic by about +0.2\,eV/H${}_2$, with a slight preference for the 
checkerboard arrangement in which the two hydrogen species maximize their mutual 
distance. This average binding energy is, however, still lower than the average 
binding energy we obtained for the $(1 \times 1)$ O${}^{\rm br}$/H${}_2^{\rm 
cus}$ phase (+0.32\,eV/H${}_2$, cf. Section IVa) which also corresponds to a 
total coverage of 1\,H${}_2$-ML.

\begin{figure}
\scalebox{0.6}{\includegraphics{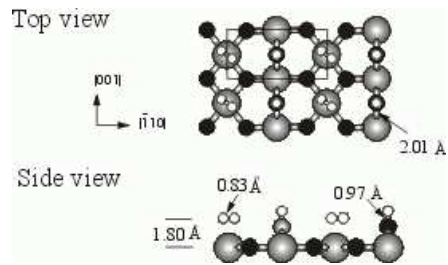}}
\caption{Top and side view of the adsorption geometry of the 3\,H-ML 
(OH)${}^{\rm br}$/H${}_2^{\rm cus}$ phase.}
\label{fig12}
\end{figure}

This leaves as interesting higher coverage phase beyond 1\,H${}_2$-ML only the
possibility to combine H${}_2^{\rm cus}$ with hydroxyl groups at the bridge 
sites. For this remaining combination even a dense $(1 \times 1)$ arrangement of 
(OH)${}^{\rm br}$/H${}_2^{\rm cus}$ at 1.5\,H${}_2$-ML total coverage turns out 
very stable with an average binding energy of +0.6\,eV/H${}_2$. This points at 
the possibility that after the aforediscussed formation of hydroxyl groups at
bridge also the molecular state at the cus sites could simultaneously be 
populated upon continued hydrogen uptake. In other words that the RuO${}_2$(110) 
surface offers the fascinating property that hydrogen may coexist both in the 
dissociated monohydride and in the non-dissociated dihydrogen state. Inspecting
the corresponding geometry shown in Fig. \ref{fig12}, the first striking effect 
of the simultaneous occupation of bridge and cus states is that the pronounced 
tilting of the hydroxyl group, cf. Fig. \ref{fig5}, has disappeared. Next, the 
H${}_2$ at the cus sites is apparently more activated compared to the situation 
discussed in Section IVA when the bridging oxygen atoms were bare: The bond 
length is increased from 0.81\,{\AA} to 0.83\,{\AA}, and the molecule resides at 
1.80\,{\AA} height, i.e. 0.05\,{\AA} closer to the surface.

\begin{figure}
\scalebox{0.6}{\includegraphics{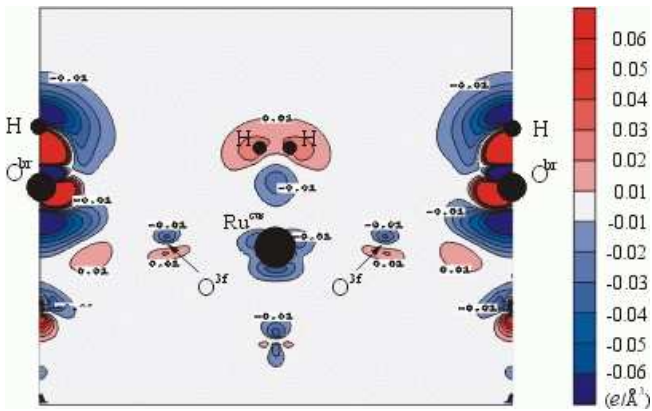}}
\caption{Difference density plot of the high coverage $( 1 \times 1)$ 
(OH)${}^{\rm br}$/H${}_2^{\rm cus}$ phase shown in Fig. \ref{fig12}. Displayed 
are the induced electron density changes in the system when hydrogen forms a 
hydroxyl group at the bridge sites. The contour plot depicts the plane normal
to the surface that cuts through a bridge site and its next-nearest neighbor cus 
site (this way passing very close a O${}^{\rm 3f}$ atom and the positions which 
are nearest to the O${}^{\rm 3f}$ are indicated by arrows). Areas of electron 
accumulation and depletion have positive and negative signs, respectively, 
contour lines are drawn at 0.01\,$e$/{\AA}$^{3}$ intervals.}
\label{fig13}
\end{figure}

Interestingly, the formation of the hydroxyl group at the bridge sites seems to 
influence the bonding properties at the neighboring cus sites, rendering the 
latter somewhat more reactive. The mechanism with which this happens is nicely 
identified on the basis of the difference density plot shown in Fig. 
\ref{fig13}. Plotted are the induced electron density variations arising in the 
(OH)${}^{\rm br}$/H${}_2^{\rm cus}$ phase when the hydroxyl group is formed. 
Apart from the obvious significant density rearrangement at the bridge sites 
themselves (which is very similar to the one shown in Fig. \ref{fig6} for the 
lower coverage (OH)${}^{\rm br}$/$-$ phase), also some variations can be 
observed at the distant cus sites: electron density is depleted around the 
Ru${}^{\rm cus}$ atoms and clearly populates the H${}_2$ $\sigma^{*}$-orbital, 
thereby weakening the molecular bond. At the same time the H${}_2$-metal bond is 
strengthened, so that the absolute binding energy is only slightly altered 
(0.23\,eV/H${}_2$ compared to the 0.32\,eV/H${}_2$ without the presence of the 
hydroxyl group, cf. Section IVA). Still, the  molecule is more activated and 
adsorbs closer to the surface.

We interpret the surface's capability to give a higher back-donation to the cus 
bond upon population of hydroxyl groups at bridge as arising from an interesting 
bond order propagation effect. Due to the newly formed hydroxyl bond the 
O${}^{\rm br}$ atoms become less bound to the underlying Ru${}^{\rm br}$ atoms, 
as reflected by the considerable increase in bond length from 1.91\,{\AA} to 
2.01\,{\AA}. Having thus lost a bit of their optimum bonding environment, the 
Ru${}^{\rm br}$ seek to fortify their remaining back-bonds, among others also to 
the directly coordinated in-plane O${}^{\rm 3f}$ atoms, cf. Fig. \ref{fig2}. The 
resulting electron density rearrangement at the latter atoms can nicely be 
discerned in Fig. \ref{fig13}, which displays a plane connecting next-nearest 
neighboring bridge and cus sites and thus passes also closely by the O${}^{\rm 
3f}$ sites. With this slightly strengthened bond to the Ru${}^{\rm br}$ atoms 
the O${}^{\rm 3f}$ atoms in turn adapt by weakening their bonds to the 
Ru${}^{\rm cus}$ atoms. Due to this propagation of bond order the latter atoms 
find themselves in a neighborhood that is less electron demanding, allowing for 
an increased back-donation into the H${}_2$-Ru${}^{\rm cus}$ bond.

While we find further reaching lateral interactions at this surface (e.g. the 
aforediscussed low coupling between hydroxyl groups at neighboring cus sites) to 
be rather small, the bonding properties at nearest-neighboring bridge and cus 
sites are thus obviously under a non-negligible mutual influence. This suggests 
that one could attempt to tune the reactivity of either site by a controlled 
population of the respective other site. Particularly for oxidation reactions at 
RuO${}_2$(110) several studies have already emphasized the key role played by 
the cus sites \cite{over00,reuter03a,reuter03c,kim00,fan01,wang01}. 
Correspondingly, modifications at the bridge sites like the present decoration 
with hydroxyl groups could have a noticeable impact on the overall catalytic
activity of the surface -- not only because of a possible site-blocking
of bridge sites, but also because of an intricate tuning of the electronic 
structure at the cus sites. So far, all reaction mechanisms discussed at this
RuO${}_2$(110) surface have been found to be initial-state dominated, i.e. the 
reaction barriers scaled with the binding energies of the adsorbed reactants 
\cite{over03,reuter03a}. The above described increased electron-donation ability 
of the cus sites upon hydroxylation of the bridge sites might therefore likely 
lead to a different reactivity of this oxide surface. Concomitantly we notice 
that a promoting effect of small amounts of hydrogen on the CO oxidation 
reaction over polycrystalline RuO${}_2$ has already been reported in a recent 
experimental study, proposing this material as a suitable candidate for the 
technologically important low-temperature CO oxidation in humid air 
\cite{zang00}. The present work identifies an increased reactivity of the cus 
sites after hydroxylation of the bridge sites at RuO${}_2$(110). Further work 
involving atom-specific Surface Science experiments on the reactivity of 
hydrogenated RuO${}_2$ surfaces is now required to check if this can be 
exploited not only for total oxidation reactions, but possibly even more 
important for partial oxidation, where a controlled tuning of the reactive sites 
might be crucial to obtain a high selectivity.

\section{Summary}

In conclusion we presented a detailed {\em ab initio} study of the energetic, 
electronic, structural and vibrational properties of hydrogen at the 
stoichiometric RuO${}_2$(110) termination. A different interaction with the two 
undercoordinated, prominent adsorption sites at this surface is found: At 
Ru${}^{\rm cus}$ only a molecular state can be stabilized, while the 
thermodynamic ground state is represented by hydroxyl groups involving the 
O${}^{\rm br}$ surface atoms. The results strongly suggest that the 
low-temperature H${}_2$ dissociation takes place via a precursor where molecular 
H$_2$ is bound at the Ru$^{\rm cus}$ site. From there a water-like dihydride 
state at O${}^{\rm br}$ can be accessed that leads to the final hydroxyl groups. 
This emerging picture of the low-temperature dissociation kinetics has been 
developed in a synergetic interplay with the experimental group of Jacobi and 
Ertl, and is in full agreement with their HREELS and TDS data \cite{wang03}, in 
particular with the peculiar blue-shifted scissor mode measured for
the water-like species at O${}^{\rm br}$.

Upon further hydrogen adsorption both the molecular H${}_2$ state at the 
Ru$^{\rm cus}$ site and the dissociated monohydride state at bridge-site surface 
oxygen atoms become populated. The formation of the hydroxyl groups is hereby
found to intricately influence the reactivity at the neighboring
cus sites, allowing for an increased back-donation further activating
the H${}_2^{\rm cus}$ bond. This modification of the bonding
properties at cus by hydrogen decoration at bridge is attributed
to a bond order propagation mechanism, possibly special to this
metallic oxide. It is argued that the resulting possibility of
fine tuning the cus site reactivity by controlled modification of
the bridge site population could be of relevance for catalytic
applications, in particular partial oxidation reactions where a
precise tuning of the bond strengths could be crucial to obtain high
selectivities.

\section*{Acknowledgments}

Q.S. is thankful for an Alexander von Humboldt fellowship. Valuable discussions
with K. Jacobi, Y. Wang, J. Wang, C.Y. Fan, and G. Ertl are gratefully 
acknowledged.

\section*{Appendix}

In order to ensure the needed numerical accuracy we particularly performed 
calculations with larger interstitital plane wave cutoffs and denser k-point 
meshes -- the latter two being the most influential parameters in the FP-LAPW 
basis set. Some of these tests are sketched in this appendix. Specifically, we 
increased the cutoffs up to 36\,Ry in several steps and doubled the k-point mesh 
to 36 k-points in the IBZ for $(1 \times 1)$ surface unit-cells. Table 
\ref{tableIII} lists the resulting bond lengths, bond angles, and binding 
energies for three characteristic phases involved in the present study: the bare 
RuO${}_2$(110) surface (O${}^{\rm br}$/$-$), a hydroxyl-group at bridge 
((OH)${}^{\rm br}$/-) and molecular hydrogen at cus (O${}^{\rm br}$/H${}^{\rm 
cus}_2$). It can be seen that the structural parameters of all three phases are 
well converged already at the lowest cutoff listed (20.25\,Ry), let alone that 
the employed k-mesh has any influence. However, the absolute binding energies 
are not yet fully converged at this low cutoff, primarily due to the need to use 
very small muffin-tin spheres as discussed in the text. From the convergence 
behavior along the sequence 
20.25\,Ry$\rightarrow$25\,Ry$\rightarrow$30.25\,Ry$\rightarrow$36\,Ry
we conclude that only at a high cutoff of 30.25\,Ry the latter quantities
are converged to within 0.1\,eV/H atom. Fortunately enough, relative energetic
differences between similar geometries involving an equal number of O and H 
atoms converge much more rapidly. This is illustrated by a comparison of the 
binding energy of the hydroxyl group at bridge either in an upright or in the 
tilted configuration (listed in Table \ref{tableIII}): at all tested cutoffs 
between 20.25\,Ry and 36\,Ry this energetic difference is 0.1\,eV/H atom, 
constant to within 0.01\,eV/H atom. Correspondingly, we employed the manageable 
cutoff of 20.25\,Ry for structural relaxations, vibrational calculations and 
when judging on the relative energetic sequence of similar structures. Only when
absolute binding energies converged to within 0.1\,eV/H atom are required,
did we run additional calculations at 30\,Ry, but with fixed geometry.

\begin{table}
\caption{\label{tableIII}
Computed bond lengths ($d$ in {\AA}), tilt angle (in ${}^{\circ}$)
and absolute binding energies ($E_b$ in eV) as a function of
interstitial plane wave cutoff and number of k-points for
three phases characteristic for the present study.}
\begin{center}
\begin{tabular}{llllllll}
$\rm O^{\rm br}/-$  &  $d_{\rm O^{\rm br}-Ru^{\rm br}}$   
&  &  &  &  & $E_b$(O)& \\ \hline 
20.25 Ry, 18 k   & 1.91  &  &  &    && 2.54        & \\
25 Ry, 18 k      & 1.90  &  &  &    && 2.46        & \\
25 Ry, 36 k      & 1.90  &  &  &    && 2.43        & \\ 
30.25 Ry, 18 k   & 1.91  &  &  &    && 2.43        & \\
36 Ry, 18 k      & 1.91  &  &  &    && 2.37        & \\
\\
${\rm (OH)^{\rm br}/- }$ & $d_{\rm O^{\rm br}-Ru^{\rm br}}$
&  &  &  $d_{\rm H^{\rm br}-O^{\rm br}}$  &  tilt &  $E_b$(H) &    \\ \hline   
20.25 Ry, 18 k      & 2.06 &   &   & 0.97 & 40 & 0.71 &       \\
25 Ry, 18 k         & 2.08 &   &   & 0.98 & 45 & 0.84 &        \\
25 Ry, 36 k         & 2.08 &   &   & 0.98 & 46 & 0.82 &        \\
30.25 Ry, 18 k      & 2.07 &   &   & 0.98 & 46 & 0.89 &        \\
36 Ry, 18 k         & 2.07 &   &   & 0.97 & 46 & 0.91 &        \\
\\
${\rm O^{\rm br}/H_2^{\rm cus}}$ &  $d_{\rm O^{\rm br}-Ru^{\rm br}}$  &
&  & $d_{\rm H-Ru^{\rm cus}}$  & $d_{\rm H-H}$  & $E_b$(H$_2$)  & \\ \hline
20.25 Ry, 18 k      & 1.92 &   &   & 1.89 & 0.81   & 0.38   &    \\
25 Ry, 18 k         & 1.92 &   &   & 1.89 & 0.81   & 0.35   &    \\
25 Ry, 36 k         & 1.92 &   &   & 1.89 & 0.81   & 0.34   &    \\
30.25 Ry, 18 k      & 1.92 &   &   & 1.89 & 0.81   & 0.32   &    \\
36 Ry, 18 k         & 1.92 &   &   & 1.89 & 0.81   & 0.30   &    \\
\end{tabular}
\end{center}
\end{table}

\end{document}